# IMAC: Impulsive-mitigation adaptive sparse channel estimation based on Gaussian-mixture model

Tingping Zhang[1,2*], Jingpei Dan[2] and Guan Gui[3]

1. School of Information Science and Engineering, Chongqing Jiaotong University, Chongqing, 400074 China
2. College of Computer Science, Chongqing University, Chongqing, 400044 China
3. Department Electronics and Information Systems, Akita Prefectural University, Akita, 015-0055 Japan

*Corresponding author. E-mails: ztp@cqjtu.edu.cn, guitohoku2009@gmail.com

**Abstract**

Broadband frequency-selective fading channels usually have the inherent sparse nature. By exploiting the sparsity, adaptive sparse channel estimation (ASCE) methods, e.g., reweighted L1-norm least mean square (RL1-LMS), could bring a performance gain if additive noise satisfying Gaussian assumption. In real communication environments, however, channel estimation performance is often deteriorated by unexpected non-Gaussian noises which include conventional Gaussian noises and impulsive interferences. To design stable communication systems, hence, it is urgent to develop advanced channel estimation methods to remove the impulsive interference and to exploit channel sparsity simultaneously. In this paper, robust impulsive-mitigation adaptive sparse channel estimation (IMAC) method is proposed for solving aforementioned technical issues. Specifically, first of all, the non-Gaussian noise model is described by Gaussian mixture model (GMM). Secondly, cost function of reweighted L1-norm penalized least absolute error standard (RL1-LAE) algorithm is constructed. Then, RL1-LAE algorithm is derived for realizing IMAC method. Finally, representative simulation results are provided to corroborate the studies.

## 1. Introduction

Broadband transmission is becoming more and more important in advanced wireless communications systems [1]–[3]. The main impairments in wireless systems are due to multipath propagation as well as additive noises. Hence, accurate channel state information (CSI) is required for coherence detection [4]. Based on assumption of Gaussian noise model, second-order statistics based least mean square (SOS-LMS) algorithm has been widely used to estimate channels due to its simplicity and robustness [5]. However, the performance of SOS-LMS is usually limited by impulsive interferences in modern communications systems [6][7]. Such impulsive noise, which results from natural or man-made electromagnetic waves, usually has a long tail distribution and violates the commonly used Gaussian noise assumption [8]. To intuitively illustrate the impulsive noise, we consider an Gaussian noise model (GMM) which is used to describe non-Gaussian noise environment [7].

For mitigating the harmful interferences, it is necessary to develop robust channel estimation algorithms in the presence of impulsive noise. Based on the assumption of dense finite impulse response (FIR), recently, several effective adaptive channel estimation algorithms have been proposed to achieve the robustness against impulsive interferences [9]–[11]. In [12], variable step-size (VSS) sign algorithm based adaptive channel estimation was proposed to achieve performance gain. In [13], an standard VSS affine projection sign algorithm (VSS-APSA) was proposed and its improved version was also proposed in [11]. However, FIR of the real wireless channel is often modeled as sparse or cluster-sparse and hence many of channel coefficients are zero [14]–[18]. Hence, these algorithms may not exploit sparse channel structure information. Indeed, some potential performance gain could be obtained if adopting advanced adaptive channel estimation methods.

To exploit channel sparsity as well as to mitigate impulsive interferences, in this paper, we propose an effective method which is termed as impulsive-mitigation adaptive sparse channel estimation (IMAC) using reweighted L1-norm penalized [19] least absolute error (RL1-LAE) algorithm. Specifically, cost function of RL1-LAE is constructed and corresponding updating equation is derived. Gaussian-mixture model (GMM) is considered to describe the impulsive noise environments. At last, several representative simulation results are provided to verify the effectiveness of the proposed IMAC method.

The rest of the paper is organized as follows. In Section 2, we introduce GMM-induced adaptive sparse system model and finds out the drawback of standard LAE. Based on GMM noise model, we propose a stable IMAC method using RL1-LAE algorithm in Section 3. In Section 4, computer simulations are given to validate the effectiveness of the propose IMAC method. Finally, Section 5 concludes the paper and brings forward the future work.

**2. System model and problem formulation**

Consider an additive noise interference channel, which is modeled by the unknown *N*-length finite impulse response (FIR) vector $\boldsymbol{w} = [w_0, w_1, \cdots, w_{N-1}]^T$ at discrete time $t$. The ideal received signal is expressed as

$$d(n) = \boldsymbol{x}^T(n)\boldsymbol{w} + z(n), \tag{1}$$

where $\boldsymbol{x}(n) = [x(n), x(n-1), \cdots, x(n-N+1)]^T$ is the input signal vector of the $N$ most recent input samples and $z(t)$ is impulsive noise which can be described by Gaussian mixture model (GMM) [7] as

$$(1-\phi)\cdot\mathcal{CN}\left(\alpha_1, \sigma_1^2\right) + \phi\cdot\mathcal{CN}\left(\alpha_2, \sigma_2^2\right) \tag{2}$$

where $\mathcal{CN}(\alpha_1, \sigma_1^2)$, $i = 1, 2$ denotes the Gaussian distributions with mean value $\alpha_i$ and variance $\sigma_i^2$, and the $\phi$ is the mixture parameter to control the impulsive noise level. According to (2), one can find that stronger impulsive noises could be described by larger $\sigma_i^2$ as well as larger $\phi$.

The objective of the adaptive channel estimation is to perform adaptive estimate of $\boldsymbol{w}(n)$ with limited complexity and memory given sequential observation $\{d(n), \boldsymbol{x}(n)\}$ in the presence of additive GMM noise $z(n)$. That is to say, the ideal observation signal $d(n)$ is given as

$$d(n) = \boldsymbol{x}^T(n)\boldsymbol{w} + z(n). \tag{3}$$

where $\boldsymbol{w}$ is an *N*-dimensional column vector of the unknown system that we wish to estimate, $z(n)$ describes the measurement noise with variance $\sigma_v^2$, and the input signal vector is $\boldsymbol{x}(n) = [x(n), x(n-1), \cdots, x(n-m+1)]^T$. We define the a prior output error vector, and the a posteriori output error vector as

$$e(n) = d(n) - \boldsymbol{x}^T(n)\boldsymbol{w}(n) \tag{4}$$

where $w(n)$ is the estimator of $w$ at iteration $n$. The cost function of standard least absolute error (LAE) is constructed as

$$G_s(n) = \|e(n)\|_1, \tag{5}$$

where $\|\ \|_1$ denotes L1-norm operator, i.e., $\|w\|_1 = \sum_{i=0}^{N-1} |w_i|$. The derivative of the cost function (5) with respect to the weight vector $w(n)$ is

$$\frac{\partial G_s(n)}{\partial w(n)} = -\mathrm{sgn}(e(n)) x(n), \tag{6}$$

where $\mathrm{sgn}(\cdot)$ denotes the sign function, i.e., $\mathrm{sgn}(x) = x/|x|$ if $x \neq 0$ and $\mathrm{sgn}(x) = 0$ if $x = 0$. Hence, the updating equation of the standard LAE is derived as

$$\begin{aligned} w(n+1) &= w(n) - \mu \frac{\partial G_s(n)}{\partial w(n)} \\ &= w(n) + \mu x(n) \mathrm{sgn}(e(n)) \end{aligned} \tag{7}$$

where $\mu$ denotes the gradient step-size. One can easy find that (7) does not exploit channel sparsity which could be exploited to improve the estimation performance. Hence, sparsity-promoting LAE algorithm is necessary to exploit the channel sparsity as well as to mitigate the impulsive noise.

## 3. Proposed RL1-LAE algorithm

### 3.1. Optimal sparse LAE algorithm with L0-norm sparse constraint

To full take advantage of channel sparsity, optimal sparse constraint function (i.e., $\ell_0$-norm) is considered for sparsity-promoting LAE algorithm for estimating channels in impulsive interference environments. With a constraint on the weight channel coefficients vectors, hence, the cost function of the optimal sparse LAE algorithm is constructed as

$$G_0(n) = \|e(n)\|_1 + \lambda_0 \|w(n)\|_0, \tag{8}$$

where $\|\ \|_0$ represents $\ell_0$-norm function and $\lambda_0$ denotes the regularization parameter to trade off the instantaneous estimation error and $\ell_0$-norm sparse penalty of $w(n)$. In the perspective of mathematical theory, adopting the $\ell_0$-norm as for sparse constraint function could exploit maximal sparse structure information in channels. However, by solving the $\ell_0$-norm is a NP-hard (non-deterministic polynomial-time hard) problem [20]. Hence, it is necessary to replace it with approximate sparse constraints so that (8) can be solvable. In the subsequent, we propose an alternative sparse adaptive filtering algorithm, i.e., RL1-LAE, to exploit the channel sparsity as well as to mitigate the impulsive interferences simultaneously.

*3.2. RL1- LAE algorithm with RL1-norm sparse constraint*

RL1 minimization for adaptive sparse channel estimation has a better performance than $\ell_1$-minimization that is usually employed in compressive sensing [19]. It is due to the fact that a properly reweighted $\ell_1$ norm approximates the $\ell_0$-norm, which actually needs to be minimized, better than $\ell_1$-norm. Hence, one approach to enforce the sparsity of the solution for the sparse LAE algorithm is to introduce the RL1 penalty term in thee cost function as RL1-LAE which considers a penalty term proportional to the RL1 of the coefficient vector. The corresponding cost function can be written as

$$G_r(n) = \|e(n)\|_1 + \lambda_r \|f(n)w(n)\|_1, \tag{9}$$

where $\lambda_r$ is the weight associated with the penalty term and elements of the $1 \times N$ row vector $f(n)$ are set to

$$[f(n)]_i = \frac{1}{\delta_r + |[w(n-1)]_i|}, \quad i = 0, 1, \cdots, N-1, \tag{10}$$

where $\delta_r$ being some positive number and hence $[f(n)]_i > 0$ for $i = 0, 1, ..., N-1$. The update equation can be derived by differentiating (9) with respect to the FIR channel vector $w(n)$. Then, the resulting update equation is:

$$\begin{aligned} w(n+1) &= w(n) + \mu \frac{\partial G_r(n)}{\partial w(n)} \\ &= w(n) + \mu x(n) \operatorname{sgn}(e(n)) - \mu \lambda_r \frac{\operatorname{sgn}(w(n))}{\delta_r + |w(n-1)|}. \end{aligned} \tag{11}$$

Please notice that in Eq. (11), since $\operatorname{sgn}(f(n)) = \mathbf{1}_{1 \times N}$, hence one can get $\operatorname{sgn}(f(n)w(n)) = \operatorname{sgn}(w(n))$. Note that although the weight vector $w(n)$ changes in every stage of this sparsity-aware RL1-LAE algorithm, it does not depend on $w(n)$, and the cost function $G_r(n)$ is convex. Therefore, the RL1 penalized RL1-LAE algorithm is guaranteed to converge to the global minimization under some conditions.

## 4. Computer Simulations

In this section, the proposed channel estimation method is evaluated in different impulsive environments with different $\sigma_2^2$ and $\phi$. For achieving average performance, $M$=1000 independent Monte-Carlo runs are adopted. The simulation setup is configured according to the typical broadband wireless communication system [3]. The signal bandwidth is 50MHz located at the central radio frequency of 2.1GHz. The maximum delay spread of $0.8\mu s$. Hence, the maximum length of channel vector $w$ is $N$=80 and its number of dominant taps is set to $K \in \{4,8\}$. To validate the effectiveness of the proposed methods, average mean square error (MSE) standard is adopted. Channel estimators are evaluated by average MSE which is defined by

$$\text{Average MSE}\{w(n)\} := \frac{1}{M}\sum_{m=1}^{M}\left\{\|w_m(n)-w_m\|_2^2 / \|w_m\|_2^2\right\}, \qquad (12)$$

where $w$ and $w(n)$ are the actual signal vector and reconstruction vector, respectively. The results are averaged over 1000 independent Monte-Carlo runs. Each dominant channel tap follows random Gaussian distribution as $\mathcal{CN}(0,\sigma_w^2)$ which is subject to $E\{\|w\|_2^2\}=1$ and their positions are randomly decided within the $w$. The received SNR is defined as $P_0/\sigma_1^2$, where $P_0$ is the received power of the pseudo-random noise (PN)-sequence for training signal. Here, please notice that the impulsive noise is often occurred suddenly. Hence, the receive SNR does not include the impulsive variance $\sigma_2^2$. Threshold parameter of RL1-LAE is set as $\delta_r = 0.01$ [21]. Detailed parameters for computer simulation are given in Tab. 1.

Tab. 1. Simulation parameters.

| Parameters | Values |
|---|---|
| Training signal | Pseudo-random Gaussian sequence |
| Channel length | $N = 80$ |
| No. of nonzero coefficients | $K \in \{4,8\}$ |
| Distribution of nonzero coefficient | Random Gaussian $\mathcal{CN}(0,1)$ |
| Received SNR for channel estimation | {5dB, 10dB, 15dB} |
| GMM noise distribution | $\alpha_1 = \alpha_2 = 0, \sigma_1^2 = 10^{(-SNR/10)}, \sigma_2^2 \in \{10,40\}$ |
| Initial step-size | $\mu = 0.01$ |
| Regularization parameters for sparse penalties | $\lambda_r = 0.0001$ |
| Threshold of the RL1-LAE | $\delta_r = 0.01$ |

In the first example, average MSE performance of the proposed method is evaluated under different GMM noise environments in Figs. 1-5. To confirm the effectiveness of the three proposed method, standard LMS method, RL1-LMS and standard LAE are considered as performance benchmarks. First of all, if the GMM noise model reduces to Gaussian case ($\phi = 0$), the proposed RL1-LAE does not earn performance gain as shown in Fig. 1. Figs. 2-5 show that the proposed RL1-LAE achieve lower MSE than RL1-LMS for $\phi = 0.2$ and different impulsive interferences which are controlled by the variance $\sigma_2^2 \in \{20,40,80\}$. Hence, the effectiveness of the proposed algorithm is confirmed in the case of two sparse channels (i.e., $K = 4$ and 8) as well as different GMM noises.

In the second example, the proposed method is evaluated in different ($\phi$) GMM interferences. It is well known that robust performance of proposed algorithms may depend highly on different impulsive interferences. Here, three kinds of impulsive interferences are considered in Fig. 6. One can find that the proposed method achieves much more performance gain than RL1-LMS under stronger impulsive interferences (bigger $\phi$). Hence, Fig. 6 shows the bigger performance is caused by the fact that RL1-LMS is deteriorated severely by stronger impulsive interferences (bigger $\phi$) while RL1-LAE is stable to mitigate the impulsive interferences.

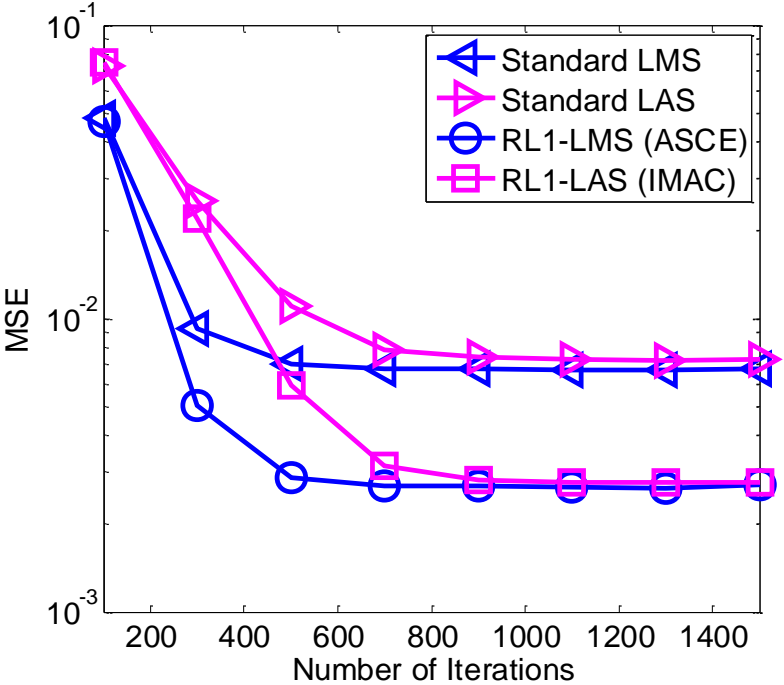

Fig. 1. Avergae MSE comparsions at T=8 and SNR=10dB with Gaussian noise model.

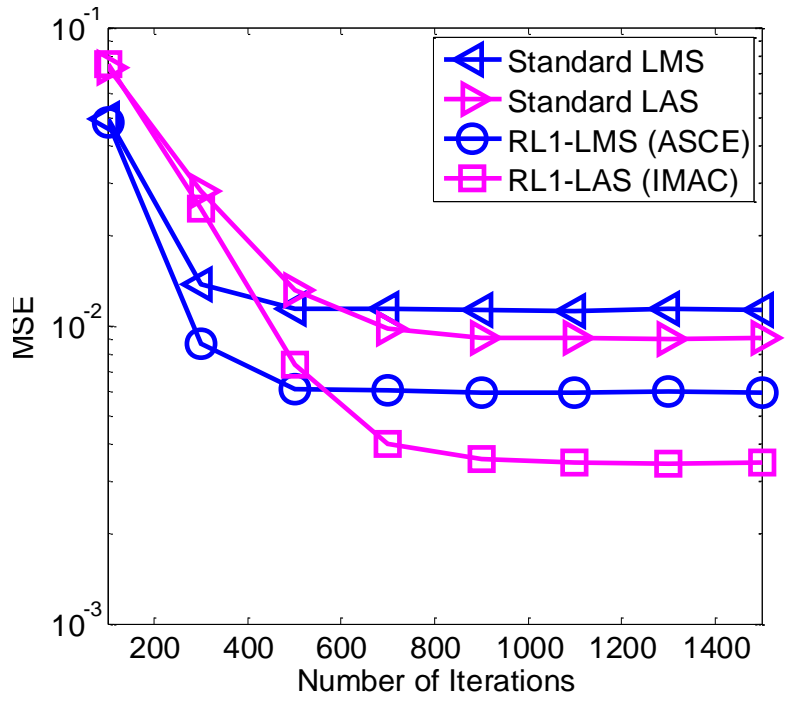

Fig. 2. Avergae MSE comparsions at T=8 and SNR=10dB with GMM ($\phi = 0.2, \sigma_n^2 = 20$).

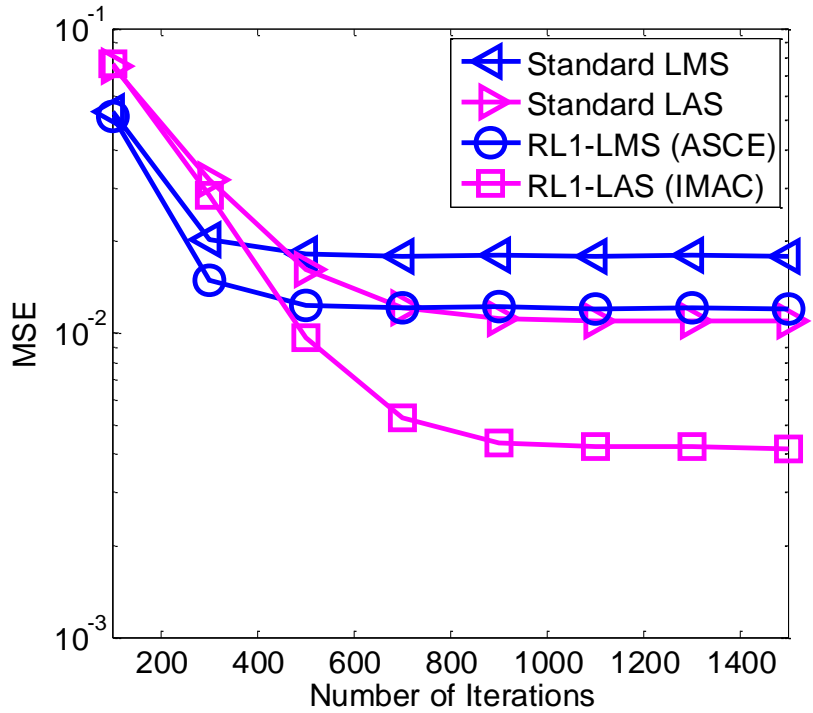

Fig. 3. Avergae MSE comparsions at T=8 and SNR=10dB with GMM ($\phi = 0.2, \sigma_n^2 = 40$).

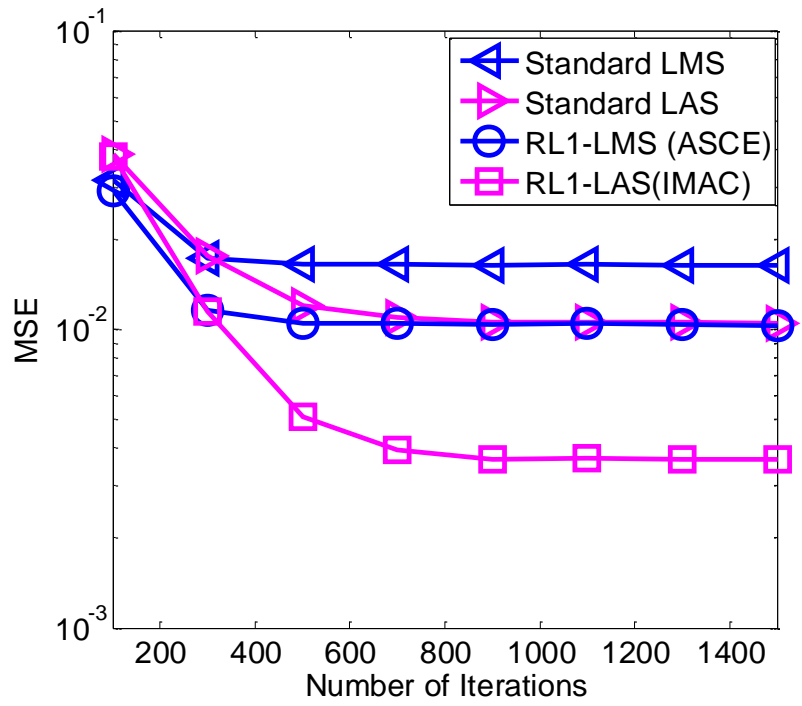

Fig. 4. Avergae MSE comparsions at T=4 and SNR=10dB with GMM ($\phi = 0.2, \sigma_n^2 = 40$).

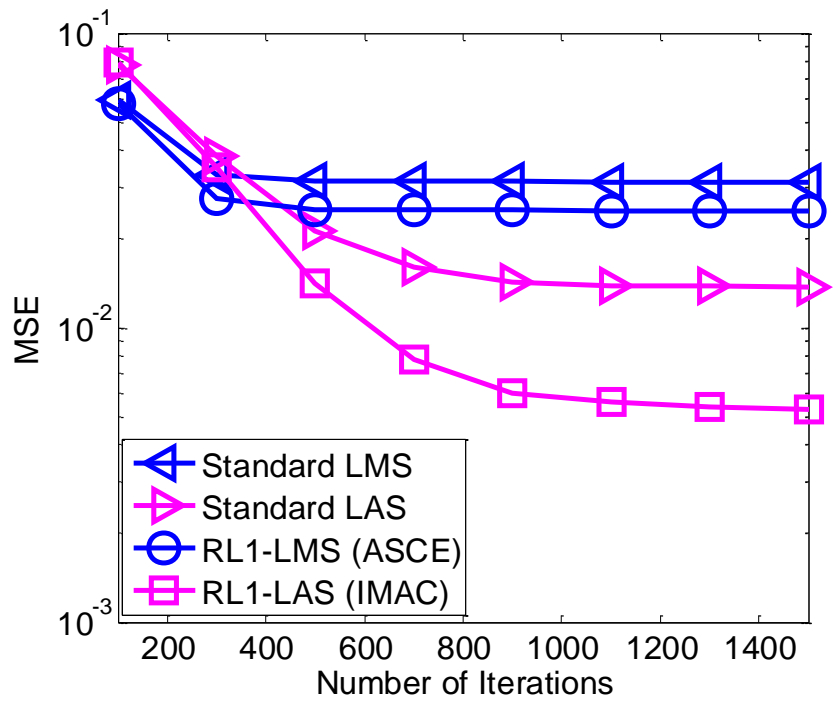

Fig. 5. Avergae MSE comparsions at T=8 and SNR=10dB with GMM ($\phi = 0.2, \sigma_n^2 = 80$).

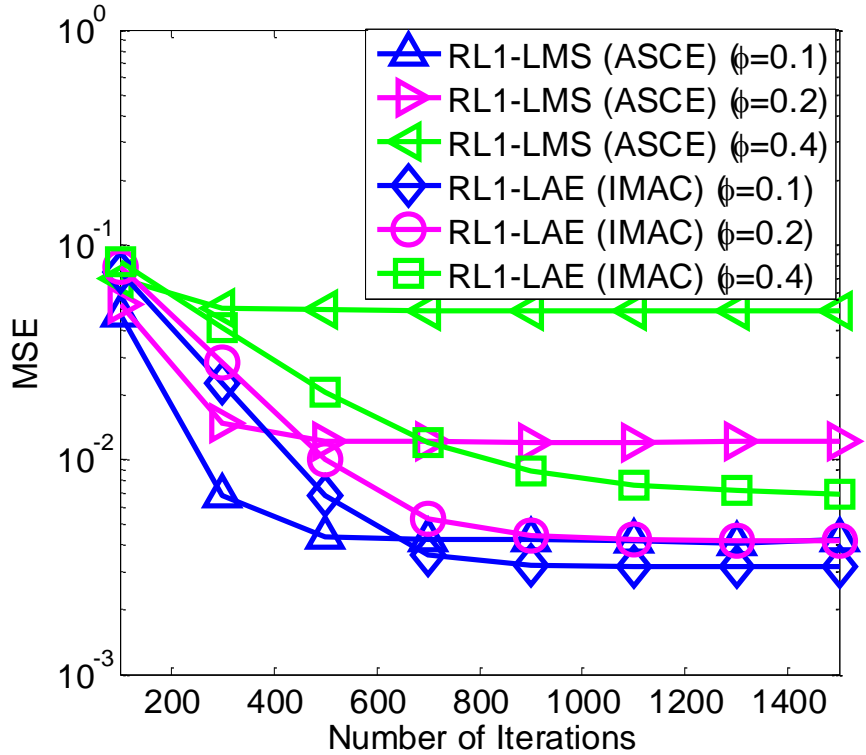

Fig. 6. Avergae MSE comparsions at T=8 and SNR=10dB with different $\phi \in \{0,1, 0.2, 0.4\}$.

## 5. Conclusions and Future Work

Gaussian noise model based conventional ASCE methods, e.g., RL1-LMS, are very sensitive to non-Gaussian noise interferences especially in the presence of very strong noise impulses. This paper proposed a stable IMAC method using RL1-LAE algorithm for mitigating the impulsive noise as well as exploiting channel sparisty. Computer simulation confirmed the proposed IMAC in different impulsive noise levels. In future work, we will test our proposed methods in different communications systems, such as underwater acoustic systems as well as power-line communication systems.